\documentclass[numberedappendix, apj, iop]{emulateapj}

\slugcomment{Submitted to ApJ}
\shorttitle{The NS Mass Distribution}
\shortauthors{K{\i}z{\i}ltan et al.}
\submitted{Submitted to the Astrophysical Journal}
\usepackage[hyperindex , breaklinks, citecolor=blue, colorlinks=true, linkcolor=linkcolor, urlcolor=true, bookmarks=true]{hyperref}
\hypersetup{colorlinks=true, linkcolor= magenta}
\usepackage{amssymb, amsmath, subfigure}
\def\be{\begin{equation}}
\def\ee{\end{equation}}
\def\bee{\begin{eqnarray}}
\def\eee{\end{eqnarray}}
\def \msun {\,\text{M}_{\odot} } 		%Msun
\def \m {\mathcal{M}}
\def\Sref#1{\S\ref{Sec:#1}}
\def\Fref#1{Figure~\ref{Fig:#1}}
\def\Tref#1{Table~\ref{Tab:#1}}
\def\Eref#1{Equation~\ref{Eq:#1}}
\newcommand{\ppd}{$\text{P}$-$\dot{\text{P}}$ }

\begin{document}

\shorttitle{The Neutron Star Mass Distribution}
\shortauthors{K{\i}z{\i}ltan et al.}
\title{The Neutron Star Mass Distribution}
\author{B\"ulent K{\i}z{\i}ltan\altaffilmark{1}, Athanasios Kottas\altaffilmark{2}  \& Stephen E. Thorsett\altaffilmark{1}}
\altaffiltext{1}{Department of Astronomy \& Astrophysics, University of California \& UCO/Lick Observatory, Santa Cruz, CA 95064}
\altaffiltext{2}{Department of Applied Mathematics and Statistics, University of California, Santa Cruz, CA 95064}

\begin{abstract}

In recent years, the number of pulsars with secure mass measurements has increased to a level that allows us to probe the underlying neutron star mass distribution in detail. We critically review radio pulsar mass measurements and present a detailed examination through which we are able to put stringent constraints on the underlying neutron star mass distribution. For the first time, we are able to analyze a sizable population of neutron star-white dwarf systems in addition to double neutron star systems with a technique that accounts for systematically different measurement errors. We find that neutron stars that have evolved through different evolutionary paths reflect distinctive signatures through dissimilar distribution peak and mass cutoff values. Neutron stars in double neutron star and neutron star-white dwarf systems show consistent respective peaks at $1.35\msun$ and $1.50\msun$ which suggest significant mass accretion ($\Delta m\approx 0.15\msun $) has occurred during the spin up phase. The width of the mass distribution implied by double neutron star systems is indicative of a tight initial mass function while the inferred mass range is significantly wider for neutron stars that have gone through recycling.  We find a mass cutoff at $2\msun$ for neutron stars with white dwarf companions which establishes a firm lower bound for the maximum neutron star mass. This rules out the majority of strange quark and soft equation of state models as viable configurations for neutron star matter. The lack of truncation close to the maximum mass cutoff suggests that the 2$\msun$ limit is set by evolutionary constraints rather than nuclear physics or general relativity, and the existence of rare super-massive neutron stars is possible. 

\end{abstract}

\section{Introduction}\label{Sec:intr}

The mass of a neutron star (NS) has been a prime focus of compact objects astrophysics since the discovery of neutrons. Soon after Chadwick's {\em Letter} on the ``Possible existence of a neutron'' (\citeyear{Chadwick:32}), heated discussions around the world started to take place on the potential implications of the discovery. In 1932, during one of these discussions in Copenhagen, Landau shared his views with Rosenfeld and Bohr where he anticipated the existence of a dense-compact star composed primarily of neutrons \citep[e.g., ][p. 242]{Shapiro:83}. The prediction was not officially announced until \citeauthor{Baade:34b} published their work where the phrase ``neutron star'' appeared in the literature for the first time \citep{Baade:34b}. Their following work explained the possible evolutionary process leading to the production of a NS and the physics that simultaneously constrains the mass and radius in more detail \citep{Baade:34a, Baade:34}. 

The ensuing discussions were primarily focused on the mass of these dense objects. In 1931, \citeauthor{Chandrasekhar:31} had already published his original work in which he calculates the upper mass limit of an ``ideal'' white dwarf as $0.91\msun$, while the following year, Landau intuitively predicted that a limiting mass should exist close to $1.5\msun$ \citep{Landau:32}. Following the works of \citeauthor{Chandrasekhar:31} and \citeauthor{Landau:32}, and using the formalism developed by \citeauthor{Tolman:39}, \citeauthor{Oppenheimer:39} predicted an upper mass limit for NSs to be $0.7$--$3.4\msun$ \citep{Tolman:39, Oppenheimer:39}.

Since then, continuing discussions on the mass range a NS can attain have spawned a vast literature \citep[e.g.,][and references therein]{Rhoades:74, Joss:76, Thorsett:99, Baumgarte:00, Schwab:10}.
 
Masses of NSs at birth are tuned by the intricate details of the astrophysical processes that drive core collapse and supernova explosions \citep{Timmes:96}. The birth mass is therefore of particular interest to those who study these nuclear processes. An earlier attempt by \cite{Finn:94} finds that NSs should predominantly fall in the $1.3$--$1.6\msun$ mass range. The most comprehensive work to date by \cite{Thorsett:99} finds that the mass distribution of observed pulsars are consistent with $M=1.38^{-0.06}_{+0.10}\msun$, a remarkably narrow mass range. The recent work of \cite{Schwab:10} on the other hand, argues that there is evidence for multi-modality in the NS birth mass distribution (for discussion, see \Sref{disc}). 

 The maximum possible mass of a NS has attracted particular attention because it delineates the low mass limit of stellar mass black holes \citep{Rhoades:74, Fryer:01}. When combined with measurements of NS radii, it also provides a distinctive insight into the structure of matter at supranuclear %($\rho_{nuc}> 3\times 10^{14}$ g~cm$^{-3}$) 
 densities \citep{Cook:94, Haensel:03, Lattimer:04, Lattimer:07}. Although more modern values theoretically predict a maximum NS mass of  $M_{max}\approx2.2$--$2.9\msun$ \citep{Bombaci:96, Kalogera:96, Heiselberg:00}, it is still unclear whether very stiff equations of states (EOSs) that stably sustain cores up to the general relativity limit ($\sim3\msun $) can exist. 

Recent observations of pulsars in the Galactic plane as well as globular clusters suggest that there may be, in fact, NSs with masses significantly higher than the canonical value of $1.4\msun$ \citep[e.g.,][]{Champion:08, Ransom:05, Freire:08, Freire:08b, Freire:08a}. NSs in X-ray binaries also show systematic deviations from the canonical mass limit \citep[e.g.,][]{van-Kerkwijk:95, Barziv:01, Quaintrell:03, van-der-Meer:05, Ozel:09, Guver:10}.

The most precise measurements of NS masses are achieved by estimating the relativistic effects to orbital motion in binary systems. The exquisite precision of these mass measurements presents also a unique means to test general relativity in the ``strong field'' regime \citep[e.g.,][]{Damour:92, Psaltis:08}. As the masses of NSs also retain information about the past value of the effective gravitational constant $G$, with the determination of the NS mass range it may be even possible to probe the potential evolution of such physical constants \citep{Thorsett:96}.

A comprehensive insight into the underlying mass distribution of NSs thus provides not only the means to study NS specific problems. It also offers diverse sets of constraints that can be as broad as the high-mass star formation history of the Galaxy \citep{Gould:00}, or as particular as the compression modulus of symmetric nuclear matter \citep{Glendenning:86, Lattimer:90}. 

The present work aims to set up a framework by which we can probe the underlying mass distributions implied by radio pulsar observations. We take a comparative approach and analyze the results that we obtain through both conventional (maximum likelihood estimation) and modern (Bayesian) statistical methods. Unlike conventional statistical methods, with a Bayesian approach it is possible to separately infer peaks, shapes and cutoff values of the distribution with appropriate uncertainty quantification. This gives us unique leverage to probe these parameters which separately trace independent astrophysical and evolutionary processes.

In order to prevent contamination of the population, which may lead to systematic deviations from the probed mass distribution, we keep the observed pulsar sample as uniform as possible. We choose mass measurements that do not have strong {\em a priori} model dependencies and therefore can be considered secure. 

In \Sref{theo} we review theoretical constraints on NS masses. We derive useful quantities such as the NS birth mass $M_{birth}$, the amount of mass expected to be transfered onto the NS primary during recycling $\Delta m_{acc}$, and the viable range of maximum mass cutoff value $M_{max}$ for NSs. The observations are reviewed in \Sref{obs}. The evolutionary paths that may produce double neutron star (DNS) and neutron star-white dwarf (NS-WD) systems are summarized in \Sref{evol}. We describe the statistical approach used to probe the underlying NS mass distribution in \Sref{esti} and detail the process through which we test the performance in \Sref{algo}. After we summarize in \Sref{summ}, the range of implications and following conclusions are discussed in \Sref{disc}. For brevity, the details of the algorithm and the analytical derivation of the numerical method for estimation are included as an appendix (\Sref{model} and \Sref{post}).

\section{Theoretical Constraints}\label{Sec:theo}

	\subsection{Birth Mass}\label{Sec:birth}

The canonical mass limit $M_{ch}\sim1.4 \msun $ is the critical mass beyond which the degenerate remnant core of a massive star or a white dwarf will lose gravitational stability and collapse into a NS. This limiting mass is an approximation which is sensitive to several nuclear, relativistic and geometric effects \citep[see][for review]{Ghosh:07, Haensel:07}. In addition to these effects, the variety of evolutionary processes that produce NSs warrant a careful treatment.

A more precise parametrization of the Chandrasekhar mass is
\be
M_{ch}=5.83\, Y_{e}^{2}
\ee
where $Y_{e}=n_{p}/(n_{p}+n_{n})$ is the electron fraction. A perfect neutron-proton equality ($n_{p}=n_{n}$) with $Y_{e}=0.50$ yields a critical mass of
\be
M_{ch}=1.457\msun.
\ee
However, we have a sufficiently good insight into the processes that affect $M_{ch}$. So, we can go beyond the idealized cases and estimate  the remnant's expected initial mass more realistically. 

The inclusion of more reasonable electron fractions ($Y_{e}<0.50$) yields smaller values for $M_{ch}$. General relativistic implications, surface boundary pressure corrections, and the reduction of pressure due to non-ideal Coulomb interactions ($e^{-}$-$e^{-}$ repulsion, ion-ion repulsion and $e^{-}$-ion attraction) at high densities all reduce the upper limit of $M_{ch}$. 

On the other hand, the electrons of the progenitor (i.e., white dwarf or the core of a massive star) material are not completely relativistic. This reduces the pressure leading to an increase in the amount of mass required to reach the gravitational potential to collapse the star. Finite entropy corrections and the effects of rotation will also enhance the stability for additional mass. These corrections, as a result, yield a higher upper limit for $M_{ch}$.

The level of impact on the birth masses due to some of these competing effects is not well constrained as the details of the processes are not well understood. An inclusion of the effects that are due to the diversity in the evolutionary processes alone requires a $\approx 20$\% correction \citep[for a detailed numerical treatment see][]{Butterworth:75} and therefore implies a broader mass range, i.e. $M_{ch}\sim1.17$--$1.75\msun$. 

The measured masses, however, are the effective gravitational masses rather than a measure of the baryonic mass content. After applying the quadratic correction term $M_{baryon}-M_{grav}\approx 0.075\, M_{grav}^{2}$, we get  
\be
M_{birth}\sim1.08\text{--}1.57\msun 
\label{eq:birth}
\ee
as a viable range for gravitational NS masses at birth.

	\subsection{Accreted Mass}\label{Sec:accr}

There is considerable evidence that at least some millisecond pulsars have evolved from a first generation of NSs which have accumulated mass and angular momentum from their evolved companion \citep{Alpar:82, Radhakrishnan:82,  Wijnands:98, Markwardt:02, Galloway:02, Galloway:05}. There is also a line of arguments that support the possibility of alternative evolutionary processes that may enrich the millisecond pulsar population \citep{Bailyn:90, Kiziltan:09}.  

Possible production channels for isolated millisecond pulsars are mergers of compact primaries or accretion induced collapse (AIC). In the case where a NS is produced via AIC, the final mass configuration of the remnant is determined by the central density of the progenitor (C-O or O-Ne white dwarf) and the speed at which the conductive deflagration propagates \citep{Woosley:92}. 

While there are uncertainties for the parameters that describe the ignition and flame propagation, a careful treatment of the physics that tune the transition of an accreting white dwarf yields a unique baryonic mass $M_{baryon}\approx 1.39\msun$ for the remnant which gives a gravitational mass of $M_{grav}\sim 1.27\msun$ for NSs produced via AIC \citep{Timmes:96}. There is indirect evidence that the occurrence rate of AICs can be significant \citep{Bailyn:90}.

The physics of these production channels are still not understood well enough to make quantitative predictions of the NS mass distribution produced via these processes. But we can estimate the mass required to spin NSs up to millisecond periods by using timescale and angular momentum arguments.

For low mass X-ray binaries (LMXBs) accreting at typical rates of $\dot{m}\sim10^{-3}\dot{\text{M}}_{\text{Edd}}$, the amount of mass accreted onto a NS in $10^{10}$yr is $\Delta m \approx0.10\msun$. We can also estimate the amount of angular momentum required to spin the accreting progenitor up to velocities that equal the Keplerian velocity at the co-rotation radius.  In order to transfer sufficient angular momentum ($L=I\times\omega$) and spin up a normal pulsar ($R\approx$12 km, $I\approx1.4\times10^{45}$ g\,cm$^{2}$) to millisecond periods, an additional mass of $\Delta m \approx0.20\msun $ is required. Hence,
\bee
\Delta m_{acc} \approx 0.10\text{--}0.20\msun 
\eee 
will be sufficient to recycle NS primaries into millisecond pulsars.

	\subsection{Maximum Mass}\label{Sec:maxi}

The mass and the composition of NSs are intricately related. One of the most important empirical clues that would lead to constraints on a wide range of physical processes is the maximum mass of NSs. For instance, secure constraints on the maximum mass provide insight into the range of viable EOSs for matter at supranuclear densities. 

A first order theoretical upper limit can be obtained by numerically integrating the Oppenheimer-Volkoff equations for a low-density EOS at the lowest energy state of the nuclei \citep{Baym:71}. This yields an extreme upper bound to the maximum mass of a NS at $M_{max}\sim 3.2\msun$ \citep{Rhoades:74}. Any compact star to stably support masses beyond this limit requires stronger short-range repulsive nuclear forces that stiffens the EOSs beyond the causal limit. For cases in which causality is not a requisite ($v \rightarrow \infty$) an upper limit still exist in general relativity $\approx 5.2\msun$ that considers uniform density spheres \citep{Shapiro:83}. However, for these cases the extremely stiff EOSs that require the sound speed to be super-luminal  ($dP/d\rho\geq c^{2}$) are considered non-physical. 

Differentially rotating NSs that can support significantly more mass than uniform rotators can be temporarily produced by binary mergers \citep{Baumgarte:00}.  While differential rotation provides excess radial stability against collapse, even for modest magnetic fields, magnetic braking and viscous forces will inevitably bring differentially rotating objects into uniform rotation \citep{Shapiro:00}. Therefore, radio pulsars can be treated as uniform rotators when calculating the maximum NS mass.

While general relativity along with the causal limit put a strict upper limit on the maximum NS mass at $\sim3.2\msun$, the lower bound is mostly determined by the still unknown EOS of matter at these densities and therefore is not well constrained. There are modern EOSs with detailed inclusions of nuclear processes such as kaon condensation and nucleon-nucleon scattering which affect the stiffness. These EOSs give a range of $1.5$--$2.2\msun$ as the lower bound for the maximum NS mass \citep{Thorsson:94, Kalogera:96}. Although these lower bounds for a maximum NS mass are implied for a variation of more realistic EOSs, it is still unclear whether any of these values are favored. Therefore, 
\bee
M_{max} \sim 1.5\text{--}3.2\msun 
\eee
can be considered a secure range for the maximum NS mass value.

\section{Observations}\label{Sec:obs}

The timing measurements of radio pulsations from NSs offer a precise means to constrain orbital parameters \citep{Manchester:77}. For systems where only five Keplerian orbital parameters (orbital period: $P_{b}$, projected semi-major axis: $x$, eccentricity: $e$, longitude and the time of periastron passage: $\omega_{0}$, $T_{0}$) are measured, individual masses of the primary ($m_{1}$) and secondary ($m_{2}$) stars, and the orbital inclination $i$ cannot be separately constrained. They remain instead related by the measured mass function $f$ which is given by
\bee
f=\frac{(m_{2}\, \text{sin}\,i)^{3}}{M^{2}}=\left(\frac{2\pi}{P_{b}}\right)^{2} x^{3}\text{T}_{\odot}^{-1}
\eee
where $M=m_{1}+m_{2}$ and masses are in solar units, the constant $\text{T}_{\odot}\equiv \text{G}\msun /\text{c}^{3}= 4.925490947\mu$s, and $x$ is measured in light seconds.

 \begin{deluxetable}{lllc} []
\tabletypesize{\footnotesize}
\small
\singlespace
\tablecolumns{4}
\tablewidth{0pt} 
\tablecaption{Double neutron star systems}
\startdata
\hline\hline
 & & &  \\ [-0.5ex]
Pulsar	&	Mass [$M_\odot$] 	& 	68\% central limits	& Refs.\tablenotemark{a}\\ [1ex]
\hline \\[-0.5ex]
	\multicolumn{4}{c}{Double neutron star binaries}\\[1ex]
\hline \\
J0737$-$3039 & & & [1] \\
pulsar A 	& 	1.3381 	&	 $\pm 0.0007$ 		& \\%[1]\\ 
pulsar B 	&	 1.2489	&	$\pm 0.0007$		&  \\%[1]\\ 
\quad total 	&	  2.58708     & 	$\pm 0.00016$			& \\ [1ex]%[1] \\ [1ex]
J1518+4904 &&& [2]\\
pulsar 	& 1.56 	& 	$+0.13/-0.44$ 		&\\% [2]\\ 
companion & 1.05	&	$+0.45/-0.11$		 & \\%[2]\\ 
\quad total 	& 2.61	 &	$\pm 0.070$		 & \\ [1ex]
B1534+12 & & & [3] \\
pulsar 		& 	1.3332 	& 	$\pm 0.0010$ 	  &\\%  [3]\\ 
companion 	& 	1.3452 	& 	$\pm 0.0010$ 	  & \\%[3]\\ 
\quad total 		&	2.678428	 & 	$\pm 0.000018$		& \\ [1ex]
J1756$-$2251 &&& [4]\\
pulsar 		& 	1.40 		& 	$+0.02/-0.03$  	 & \\% [4]\\
companion 	& 	1.18 		& 	$+0.03/-0.02$ 	  & \\%[4]\\
\quad total 		& 	2.574		&	$\pm 0.003$	 & \\ [1ex]
J1811$-$1736 &&& [5, 6]\\
pulsar 		& 	1.56		& 	$+0.24/-0.45$		&  \\
companion 	& 	1.12	& 	$+0.47/-0.13$			&   \\
\quad total 	& 	2.57		 &	$\pm 0.10$ &  \\ [1ex]
J1829+2456 &&& [7]\\
pulsar 		& 	1.20	& 	$+0.12/-0.46$		&   \\
companion 	& 	1.40	& 	$+0.46/-0.12$		&  \\
\quad total 	& 	2.59		 & 	$\pm 0.02$		&   \\ [1ex]
J1906+0746 &&& [8, 9]  \\
pulsar 		& 	1.248 	& 	$\pm 0.018$	 	&   \\
companion 	& 	1.365		& 	$\pm 0.018$		&  \\
\quad total 		&	2.61 		& 	$\pm 0.02$			& \\ [1ex]
B1913+16 &&& [10, 11]\\
pulsar 		& 	1.4398	& $\pm 0.002$ 	& \\%  [8] \\ 
companion 	&	1.3886 	& $\pm 0.002$ 	&  \\ %[1ex]% [8] \\  [1ex]
\quad total 	 	& 	2.82843	&  $\pm 0.0002$ &\\ [1ex]
B2127+11C &&& [12]\\
pulsar		&	 1.358 	&	 $\pm 0.010$ 	  &\\% [9]\\ 
companion 	&	 1.354 	&	 $\pm 0.010$	  & \\%[9]\\
\quad total	 	& 	2.71279	&	$\pm 0.00013$	 & \\ [1ex]
\hline
\enddata
\tablenotetext{a}{References:
%*: This work,
1: \cite{Kramer:06},
2: \cite{Thorsett:99},
3: \cite{Stairs:02},
4: \cite{Faulkner:05},
5: \cite{Stairs:06},
6: \cite{Corongiu:07},
7: \cite{Champion:05},
8: \cite{Kasian:08},
9: \cite{Lorimer:06},
10: \cite{Weisberg:10},
11: \cite{Taylor:92},
12:  \cite{Jacoby:06}
}
\label{Tab:doub}
\end{deluxetable}
 %Tab:DNS  (1)

For some binary systems, the timing residuals cannot be modeled with only Keplerian parameters when the effects of general relativity are measurable. In these cases, the gravitational influence can be parametrized as five potentially measurable post-Keplerian (PK) parameters which have similar interpretations \citep{Taylor:92}; (1) $\dot{\omega}$: advance of periastron (2) $\dot{P}_{b}$: orbital period decay  (3) $\gamma$: time dilation-gravitational redshift (4) $r$: range of Shapiro delay (5)  $s$: shape of Shapiro delay, where these are described by
\bee
\dot{\omega} & = & 3\left(\frac{P_{b}}{2\pi}\right)^{-5/3}\left(\text{T}_\odot M\right)^{2/3}\left(1-e^2\right)^{-1} ,
\eee
\bee
\dot{P}_b & = & -\frac{192\pi}{5}\left(\frac{P_{b}}{2\pi}\right)^{-5/3}  \left(1+\frac{73}{24}e^2+\frac{37}{96}e^4\right)\times \nonumber\\
 & & (1-e^2)^{-7/2}\,\text{T}_\odot^{5/3}\,m_1m_2\,M^{-1/3} ,
 \eee
\bee
\gamma & = & e\left(\frac{P_{b}}{2\pi}\right)^{1/3}\text{T}^{2/3}_\odot M^{-4/3}m_2\left(m_1+2m_2\right) ,
\eee
\bee
r & = & \text{T}_\odot m_2 ,
\eee
\bee
s & = & x\left(\frac{P_{b}}{2\pi}\right)^{-2/3}\text{T}_\odot^{-1/3}M^{2/3}\,m_2^{-1}.
\label{Eq:PK}
\eee
A comprehensive review of the observational techniques and measurements can be found in \cite{Lorimer:04} and \cite{Stairs:06}.

\begin{deluxetable}{lllr}[]
\tabletypesize{\footnotesize}
\small
\singlespace
\tablecolumns{4}
\tablewidth{0pt} 
\tablecaption{Neutron star - white dwarf binary systems}
\startdata
\hline\hline
 & & &  \\ [-0.5ex]
Pulsar	&	Mass [$M_\odot$] 	& 	68\% central limits & Refs.\tablenotemark{a}	\\ [1ex]
\hline \\ [-0.5ex]
\multicolumn{4}{c}{Neutron star - white dwarf binaries}\\ [1.ex]
\hline \\
J0437$-$4715	&	1.76 		&	$\pm 0.20$				& [1] \\
J0621+1002	&	1.70		&	$+0.10/-0.17$				&[2]	\\	
J0751+1807	&	1.26		&	$\pm 0.14$			&[2]\\ 
J1012+5307 	& 	1.64 		& 	$\pm 0.22$ 	 	& [3]\\%, Lange:01} \\ 
J1141$-$6545	&	1.27		&	$\pm 0.01$				&[4]	\\
J1614$-$2230		&	1.97		&	$\pm 0.04$			&	 [5] \\ 	 
J1713+0747 	& 	1.53		& 	$+0.08/-0.06$ 	& [6]\\ 
J1802$-$2124	&	1.24 		 &	 $\pm 0.11$ 		&[7]\\ 
B1855+09 		& 	1.57 		& 	$+0.12/-0.11$ 		& [8] \\ 
J1909$-$3744	&	1.438	&	$\pm 0.024$			&  [9]	\\
B2303+46		&	1.38		&	$+0.06/-0.10$			&[10]\\ [1ex]
\hline\\ [0.5ex]
\multicolumn{4}{c}{Neutron stars in globular clusters}\\ [1.ex]
\hline\\
J0024$-$7204H 		&	1.48		&	$+0.03/-0.06$		& [*] 	\\
J0514$-$4002A 	&     1.49		& $+0.04/-0.27$  	& [*]	\\
B1516+02B 	& 2.10		&	$\pm 0.19$			&  [*]	\\
J1748$-$2446I	&	1.91		&	$+0.02/-0.10$	& [*]	 \\
J1748$-$2446J	&	1.79		&	$+0.02/-0.10$	& [*] \\
B1802$-$07  	&	1.26		 & 	$+0.08/-0.17$			& [10] \\
B1911$-$5958A	&	1.40 		& 	$+0.16/-0.10$  		&[11] \\ [1.ex]
\hline
\enddata
\tablenotetext{a}{References: 
*: This work; Freire (personal communication), 
1: \cite{Verbiest:08},
2: \cite{Nice:08},
3: \cite{Callanan:98},
4: \cite{Bhat:08},
5: \cite{Demorest:10}
6: \cite{Splaver:05},
7: \cite{Ferdman:10},
8: \cite{Nice:03},
9: \cite{Jacoby:05},
10:  \cite{Thorsett:99},
11: \cite{Bassa:06}%,
}
\label{Tab:nswd}
\end{deluxetable}
 %Tab: NS-WD (2)

In systems where at least two PK parameters can be measured, $m_{1}$ and $m_{2}$ may be individually determined. In rare cases, more than two PK parameters are measurable. These over-constrained systems present a unique means to test for consistent strong-field gravitational theories \citep{Taylor:89}. 

In \Tref{doub} and \Tref{nswd} we compile a comprehensive list of well-measured masses. We include the mass estimates along with the 68\% confidence limits which are plotted on \Fref{psr}. 

\begin{figure}[]
\includegraphics[width=0.49 \textwidth,angle=0, trim=2.cm .1cm .1cm .1cm, clip=true]{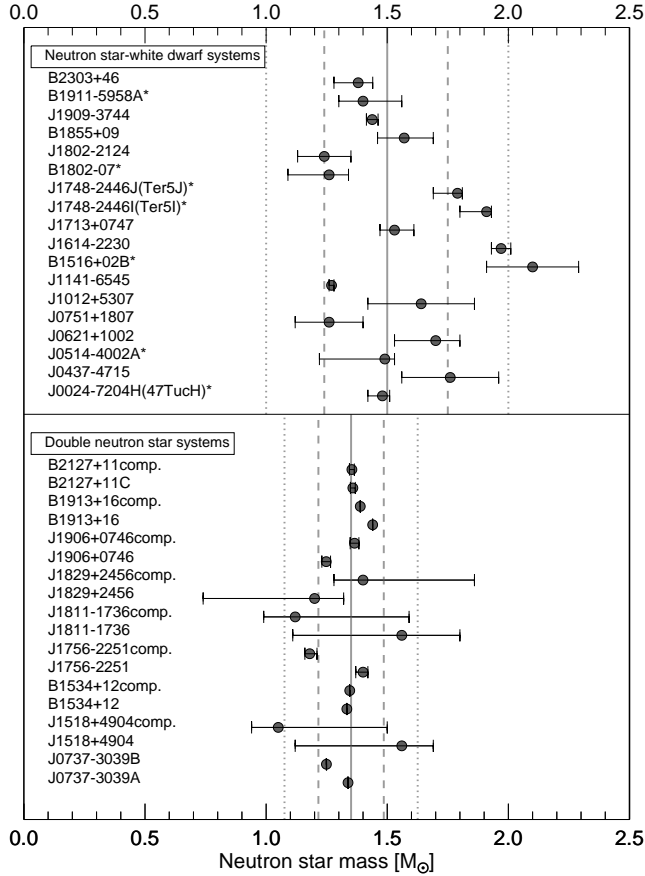} 	
\caption{Measured masses of radio pulsars. All error bars indicate the central 68\% confidence limits. Vertical solid lines are the peak values of the underlying mass distribution for DNS ($m=1.35\,\msun$) and NS-WD ($m=1.50\,\msun$) systems. The dashed and dotted vertical lines show the central 68\% and 95\% predictive probability intervals of the underlying mass distribution shown in \Fref{post}. ``$\star$'' points to pulsars found in globular clusters.}
\label{Fig:psr}   
\end{figure} % Fig:psr

We aim to prevent  possible contamination of the sample with sub-populations which may have gone through different and not well understood evolutionary paths (e.g., isolated NSs). Even for the better constrained formation processes that lead to the production of DNS and NS-WD systems, theoretical models estimating the final NS masses are tentative.

\section{Evolution}\label{Sec:evol}

While the production channels proposed for DNS and NS-WD systems are very diverse, the precise orbital parameters derived from radio observations allow us to probe the viability of these models, and hence extract constraints on their mass evolution. We therefore limit our analysis to DNS and NS-WD systems for which testable evolutionary models and reliable mass measurements exist.

	\subsection{Double Neutron Star Systems}
	
Several scenarios have been suggested for the formation of DNS systems. Each NS in DNS systems is believed to originate from massive main sequence stars with masses that exceed 8$\msun$. While the formation sequence and processes are not well understood, the first formed NS produced by the more massive primary may initially accumulate additional mass through wind accretion when the less massive secondary continues to undergo nuclear evolution during the early phases of the red giant branch. In the standard scenario, the system enters a high-mass X-ray binary (HMXB) phase in which unstable mass transfer leads to a common envelope evolution. The first formed NS is then expected to accumulate additional mass during this phase and form pulsars such as B1913+16. A double-core model has also been suggested in which a He and a CO star evolves through a common envelope phase following the initial Roche-lobe overflow \citep{Podsiadlowski:05}.  After the common envelope and a second phase of mass transfer, DNS systems such as J0737$-$3039 can be produced following two consecutive SN explosions. Alternative evolutionary scenarios in which the progenitor of the secondary (less massive) pulsar is a main sequence star with a mass less than 2$\msun$ have also been proposed as a viable production channel \citep{Stairs:06a}.

	\subsection{Neutron Star-White Dwarf Systems}

The evolutionary paths that may lead to the formation of NS-WD systems include possible episodes of accretion through wind, disk or a common envelope. Secular disk accretion is generally accepted as the dominant process of mass transfer for long-period NS-WD systems with low-mass white dwarf companions (e.g., PSR J1713+0747). On the other hand, NSs with more massive white dwarf secondaries in close orbit systems are expected to go either primarily through a common envelope phase (e.g., PSR J1157+5112) or Roche-lobe overflow followed by mass transfer through common envelope (e.g., PSR J1141$-$6545) \citep{Stairs:04}. It's been also suggested that it may be possible to produce NS-WD binary systems with orbital parameters that resemble PSR J2145$-$0750 if the donor stars fill their Roche-lobe on the asymptotic giant branch \citep{van-den-Heuvel:94}.

\section{Estimating the Underlying Mass Distribution}\label{Sec:esti}

Recent advances in statistical methods have reached a level which allows us to extract information from sparse data with unprecedented detail. Generally, there is an inverse correlation between the level of sophistication of the model and the confidence of the prediction. By dynamically measuring the performance (see \Sref{algo}) one can choose an optimal level of detail to be implemented into the model.

It can be clearly argued why modeling the underlying NS mass distribution as a single homogenous population is over-simplistic. There is no compelling line of reasoning that would require a single coherent (unimodal) mass distribution for NSs that we know have dissimilar evolutionary histories and possibly different production channels \citep[e.g., see][]{Podsiadlowski:04}. In fact, there is an increasing number of measurements that show clear signatures for masses that deviate from the canonical value of 1.4$\msun$. For instance, recent findings of \cite{van-Kerkwijk:10} imply that the mass for PSR B1957+20 may be as high as 2.4$\pm$0.12$\msun$. Many of NSs in globular clusters also show systematically higher masses \citep[see][]{Freire:08a}. Therefore, it is necessary to infer the implied mass distributions separately for different sub-populations (DNS vs. NS-WD). As we show in \Sref{algo}, an extensively tested and calibrated numerical method can then be used to test whether the implied masses belong to the same distribution. We argue that with the number of secure mass measurements available (\Tref{doub} and \Tref{nswd}), clear signatures should be manifest in the inferred underlying mass distributions if appropriate statistical techniques are utilized. Since we still operate in the sparse data regime, it is useful, if not necessary, to use Bayesian inference methods.

For the range of calculations we use mass measurements obtained directly from pulsar timing. The methods used for estimating NS masses other than radio timing, have intrinsically different systematics, and therefore require a more careful treatment when assessing the implied NS mass distribution. The inclusion of mass estimates of NSs in X-ray binaries along with these more secure measurements would potentially perturb the homogeneity of the sample and the coherence of the inference. 

For an all inclusive assessment of NS masses, more sophisticated hierarchical inference methods may be required. For sparse data, a proper statistical treatment of different systematic effects and {\em a priori} assumptions is not trivial. Also, the expected loss in precision may outweigh the gain obtained from a more detailed approach. Without properly tested and calibrated tools, further inclusion of NSs whose masses are not measured by pulsar timing may just contaminate the sample and can therefore be misleading \citep[e.g., see][]{Steiner:10}.

	\subsection{Statistical Model}\label{Sec:mode}

Here, we present the statistical model used for estimating the NS mass distribution. The approach is based on a formulation that incorporates
measurement errors of NS mass estimates. Specifically, we perform our calculations for mass distributions characterized by
\bee
m_{i} = \mathcal{M}_{i}+w_{i},  \,\,\,  i=1,...,n
\eee
where $m$ is the pulsar mass estimate and $\mathcal{M}$ is the NS mass with an associated $w$ error. We assume a normal NS mass distribution,
$N(\mathcal{M};\mu,\sigma^{2})$, with mean $\mu$ and variance $\sigma^{2}$. The errors ($w_{i}$), associated with the pulsar mass estimates ($m_{i}$) are assumed to arise from normal distributions $N(0,S_{i}^{2})$. The observation specific error variances ($S_{i}^{2}$) are obtained from the error bands of pulsar observations (i.e., \Tref{doub} and \Tref{nswd}). Assuming independence between the normal distributions for $\mathcal{M}$ and $w$, the probability model described above yields a 
\bee
N(m;\mu,\sigma^{2}+S^{2})
\label{Eq:func}
\eee
distribution for the NS mass estimates. 

Therefore, the likelihood function for the NS mass distribution parameters $(\mu,\sigma^{2})$ is given by 
\bee
\lefteqn{\mathcal{L}(\mu,\sigma^{2};\text{data})=} \nonumber \\
& & \prod^{n}_{i=1} \left [2\pi (\sigma^{2}+S_{i}^{2}) \right]^{-1/2} 
\, e^{-\frac{(m_{i}-\mu)^{2}}{2(\sigma^{2}+S_{i}^{2})}}
\label{Eq:like}
\eee
(for derivation see \Sref{model}). Here, the $\text{data}$ vector comprises the observed mass estimates $m_{i}$, and error variances
$S_{i}^{2}$, which are computed using the estimated error bars for each $m_{i}$, $i=1,...,n$. Numerical maximization of the likelihood function yields maximum likelihood estimates for the Gaussian mean $\mu$ and half-width $\sigma$ implied by pulsar observations. 

However, the general non-standard fashion of how $\sigma^{2}$ enters the expression for the likelihood function, the uncertainty quantification for the point estimates of $\mu$ and $\sigma$, and the subsequent effect on NS mass density estimates would require asymptotic results for likelihood-based confidence intervals. Given that the likelihood approach relies on large sample sizes for uncertainty estimates, this can be especially problematic where the number of mass estimates from DNS and NS-WD systems are small. 

We thus employ a Bayesian approach to modeling and inference of the NS mass distribution. Under the Bayesian model formulation, the likelihood function $\mathcal{L}(\mu,\sigma^{2};\text{data})$ is combined with (independent) prior distributions $\pi(\mu)$ and  $\pi(\sigma^{2})$ for $\mu$ and $\sigma^{2}$ to obtain the posterior  distribution for the model parameters, given the data,
\bee
p(\mu,\sigma^{2} \mid \text{data}) = C^{-1} \pi(\mu) \pi(\sigma^{2})
\mathcal{L}(\mu,\sigma^{2};\text{data}).
\eee
We work with a normal prior for $\mu$ with mean $a$ and variance $b^{2}$, and an inverse-gamma prior for $\sigma^{2}$ with mean $d/(c-1)$ for $c > 1$ (see \Sref{model} for details). The normalizing constant of the posterior distribution involves the marginal likelihood for the data, that is, 
\bee
C= \int \pi(\mu) \pi(\sigma^{2}) \mathcal{L}(\mu,\sigma^{2};\text{data})
\, d\mu d\sigma^{2}.
\eee

The posterior density is not available in closed form, since the integral for the normalizing constant cannot be  analytically evaluated. We therefore resort to a Markov chain Monte Carlo (MCMC) approach to sampling from the posterior distribution \citep[see][]{Gelman:03}. MCMC posterior inference is based on simulation from a Markov chain whose stationary distribution is given by the posterior distribution for the model parameters. As detailed in \Sref{model}, the MCMC algorithm samples dynamically from the posterior full conditional distributions for $\mu$ and $\sigma^{2}$. The former is a normal distribution and hence readily sampled; the latter is not of a standard form and thus a Metropolis-Hastings (M-H) step is used to sample from the conditional posterior distribution of $\sigma^{2}$. The resulting posterior samples for $(\mu,\sigma^{2})$ can be used for full and exact inference for the model parameters $\mu$ and $\sigma^{2}$. More importantly, the posteriors for $(\mu,\sigma^{2})$ are used to infer the NS mass distribution.

\begin{figure}[!]
\includegraphics[width=0.36 \textwidth,angle=90,trim=.5cm .5cm .cm .5cm, clip=true]{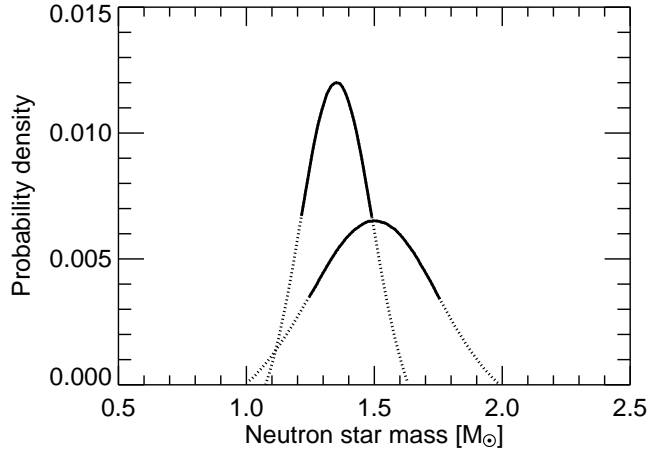} 	
\caption{Posterior predictive density estimates for the neutron star mass distribution. DNS and NS-WD systems have respective peaks at 1.35$\,\msun$ and 1.50$\,\msun$. Probability densities are normalized to show the 95\% posterior probability range. The solid parts of the curves show the central 68\% probability range which correspond to $1.35\pm0.13\,\msun$ and $1.50\pm0.25\,\msun$ for the DNS and NS-WD system, respectively.
}
\label{Fig:post}   
\end{figure} %Fig:post

In a Bayesian approach, the posterior predictive density, denoted by $\mathcal{P}(\m_{0} \mid \text{data})$, provides the estimate for the density of the NS mass distribution. Formally, $\mathcal{P}(\m_{0} \mid \text{data})$ is the distribution for a ``new'' unobserved pulsar with unknown mass $\mathcal{\m}_{0}$, which we seek to estimate (predict) given the observed data. Following the derivation in \Sref{post}, the posterior predictive density can be derived as 
\bee
\lefteqn{\mathcal{P}(\m_{0} \mid \text{data}) =}  \nonumber\\
 & &\int\int N(\m_{0};\mu,\sigma^{2})\, 
p(\mu,\sigma^{2} \mid \text{data}) \, d\mu \, d\sigma^{2}
\label{Eq:prob} 
\eee
and can thus be readily estimated using the MCMC samples from the posterior distribution $p(\mu,\sigma^{2} \mid \text{data})$. Figure~\ref{Fig:post} shows the inferred posterior predictive NS mass densities for the DNS and NS-WD systems. The 68\% and 95\% predictive probability intervals inferred from \Fref{post} are projected onto \Fref{psr} as vertical lines.

	\subsection{Comparison with Likelihood Estimation}\label{Sec:stat}

We have taken a comparative approach to probe the underlying mass distribution of NSs. In particular, in order to assess the sensitivity of the results to the technique that is used for inference, we studied the shape of the posterior density $p(\mu,\sigma^{2} \mid \text{data})$ under different priors, $\pi(\mu)$ and $\pi(\sigma^{2})$, relative to the likelihood function for $(\mu,\sigma^{2})$. 

The likelihood surface contours for the Gaussian mean $\mu$ and half-width $\sigma$ are shown in \Fref{conf}(a) for DNS and NS-WD systems. \Fref{conf}(b) plots the corresponding contours of the
posterior density under priors (2) and (3) in \Fref{priors} for DNS and NS-WD systems, respectively.

It is a good indication for the model and method acceptability if the compared parameter estimates are not drastically different. Indeed, with regard to the implied estimates for $\mu$ and $\sigma$, the difference between the maximum 
likelihood and the Bayesian approach appears practically insignificant. Section \ref{Sec:algo} includes a more detailed study of the effect of the prior choice on posterior inference results under the proposed Bayesian model.

\begin{figure}[]
\includegraphics[width=0.39 \textwidth,angle=90,trim=.5cm .5cm .cm 1.8cm, clip=true]{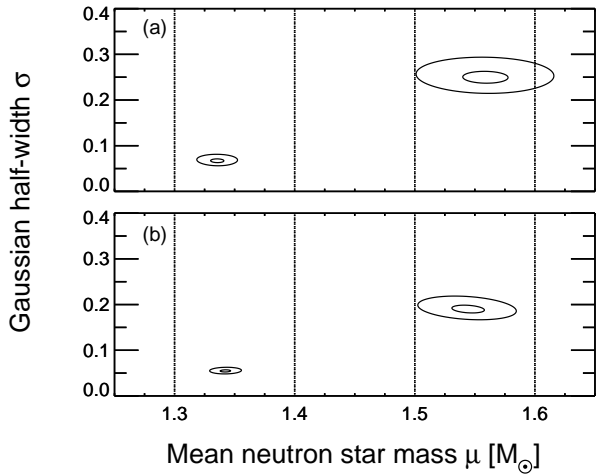} 	
\caption{Likelihood surfaces (a) and posterior densities (b) of model parameters $\mu$ and $\sigma$ for the NS mass distribution. In each panel, the tight contours on the left and wider contours on the right correspond to the data from DNS and NS-WD systems, respectively. 
}
\label{Fig:conf}   
\end{figure}
 %Fig:conf

\section{Prior Choice and Algorithm Performance}\label{Sec:algo}

	\subsection{Approach to Prior Choice}\label{Sec:like}

In a Bayesian framework, weakly informative priors are desirable for sparse data. Non-informative priors can have strong and undesirable implications that lead to artificially biased inferences. Weakly informative priors, however, will let the data tune the posterior more optimally while being strong enough to exclude various ``unphysical'' possibilities \citep{Gelman:03, Robert:07}.

In order to prevent introducing either a strongly informative bias or loss of information, we test a vast range of priors on simulated samples. The most natural way to study the effect of the prior choice is through the prior predictive density, denoted by $\mathcal{P}(\m_{0})$, which yields the prior estimate of the density for the NS mass distribution, that is, before data is used. Analogously to the expression for the posterior predictive density (see Equation \ref{Eq:prob}), the prior predictive density is defined by 
\[
\mathcal{P}(\m_{0}) = \int\int N(\m_{0};\mu,\sigma^{2}) \, 
\pi(\mu) \pi(\sigma^{2}) \, d\mu \, d\sigma^{2}.
\]
Hence, $\mathcal{P}(\m_{0})$ can be estimated by Monte Carlo integration of the $N(\m_{0};\mu,\sigma^{2})$ density, using samples from the prior distributions. A series of results are produced for priors (for $\mu$ and $\sigma^{2}$) that 
we incrementally tune to obtain prior predictive densities with dispersions that range from fairly non-informative (practically flat) to very concentrated shapes. A sample range is shown in \Fref{priors} for four combinations of hyper-parameters $(a,b)$ and $(c,d)$ that define the prior for $\mu$ and $\sigma^{2}$, respectively. We then use these priors to test the effects of the choice on the posterior distribution.

\begin{figure}
		\includegraphics[width=0.36 \textwidth,angle=90, trim=.5cm .5cm .5cm .7cm, clip=true]{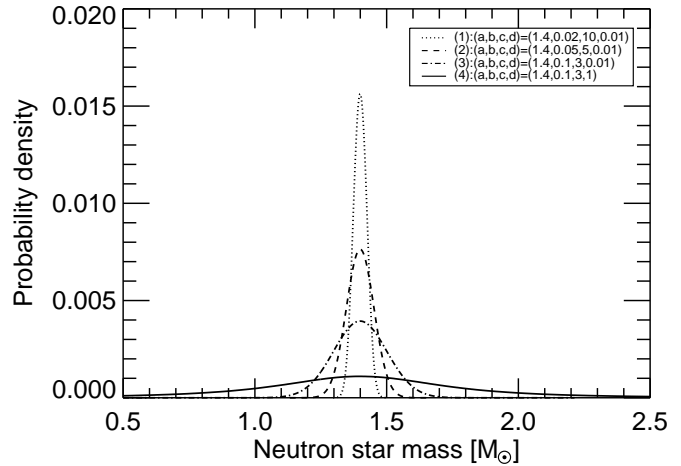} 	
\caption{Plot of the prior predictive NS mass densities under four different prior choices tested for performance. 
Each prior is defined by hyper-parameters (a,b,c,d) (see \Eref{pri1} and \Eref{pri2}). }
\label{Fig:priors}   
\end{figure}
 % Fig:priors

\begin{figure*}[!th]
\subfigure[Neutron star - white dwarf systems]{\includegraphics[width=0.38 \textwidth,angle=90,trim=.5cm .5cm .5cm 1.9cm, clip=true]{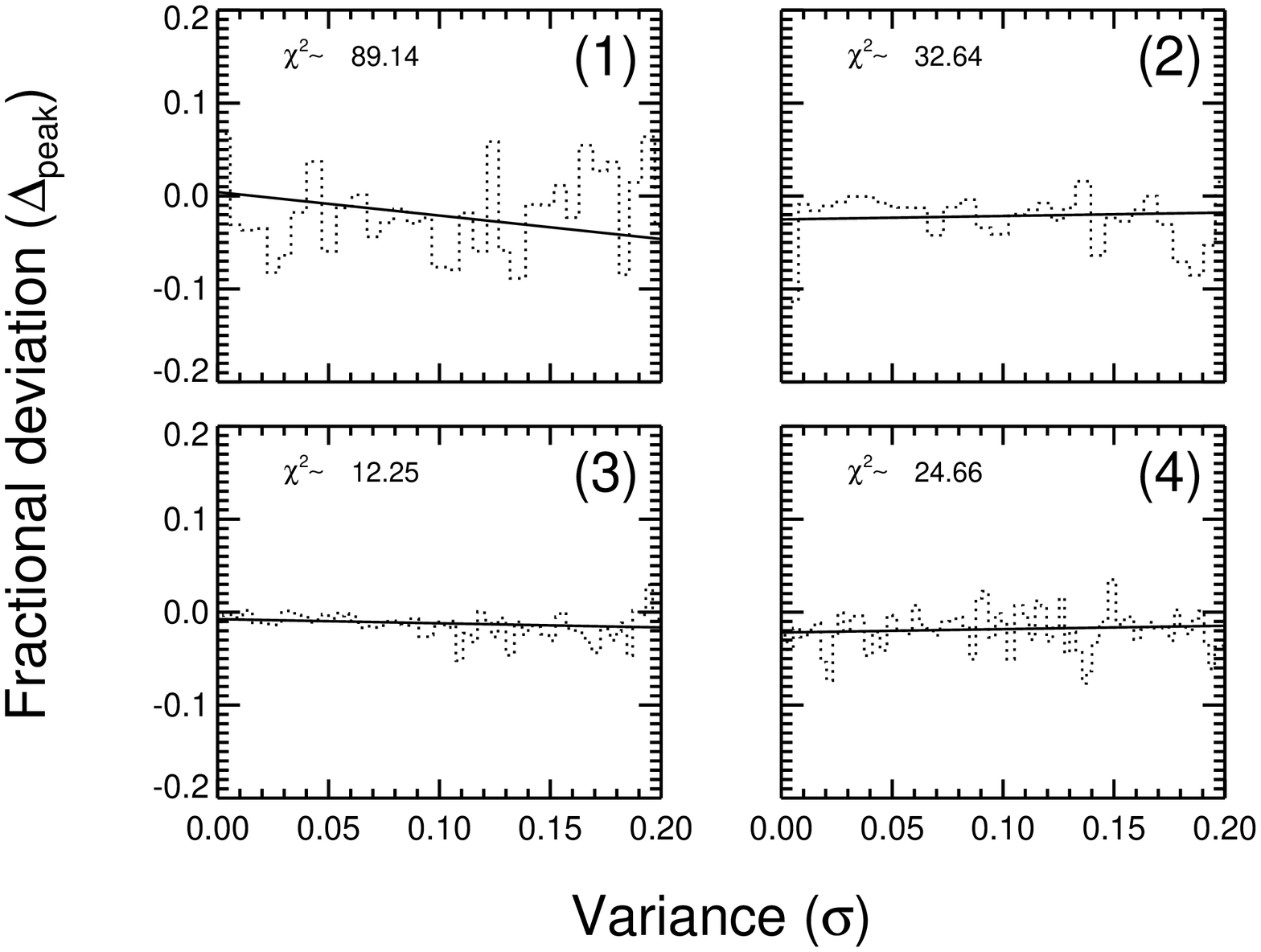} 
	\label{Fig:perfW}}
\subfigure[Double neutron stars systems]{\includegraphics[width=0.38 \textwidth,angle=90,trim=.5cm .5cm .5cm 1.9cm, clip=true]{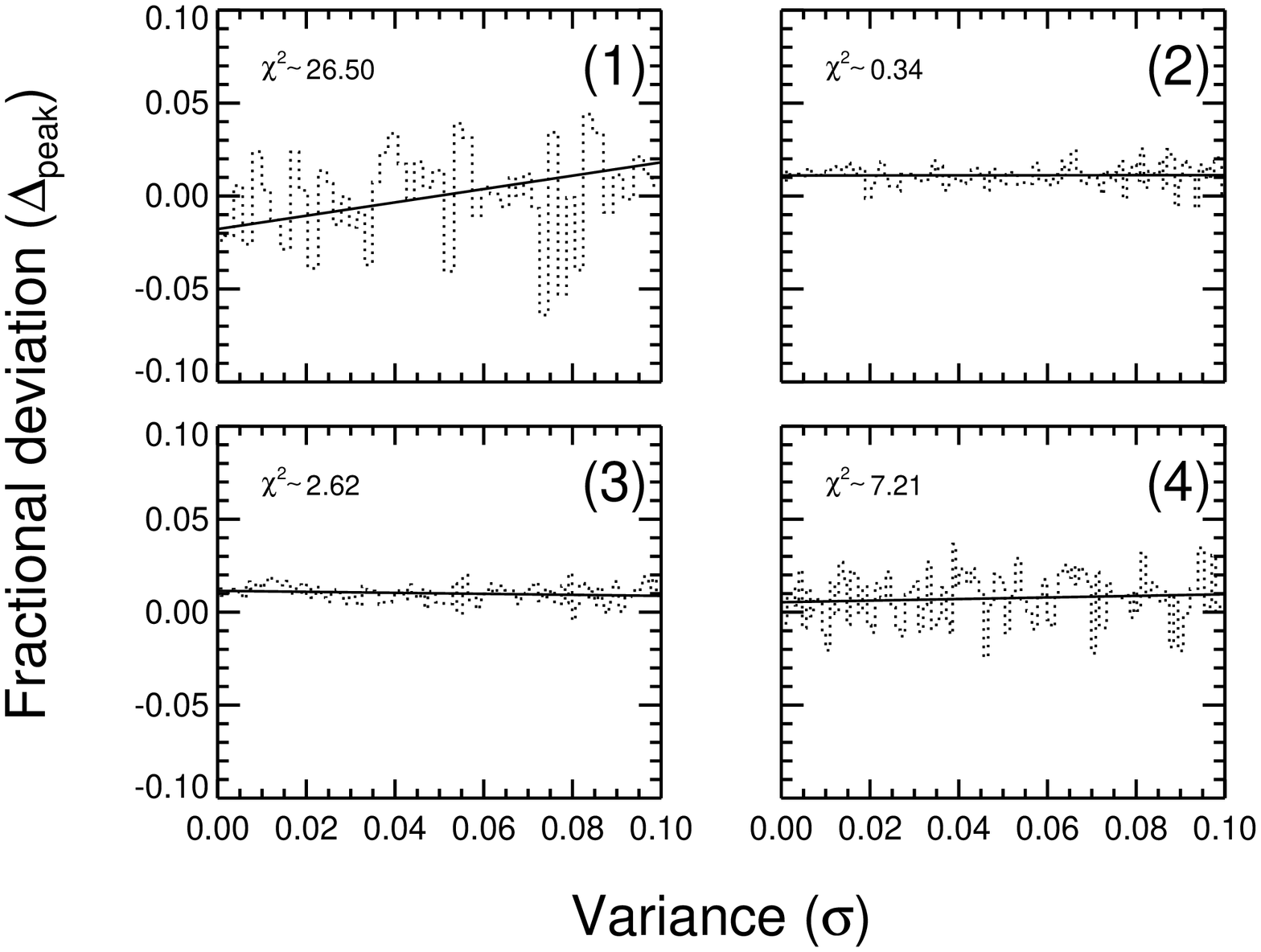}
	\label{Fig:perfD}}
\caption{Performance of the inference algorithm for (a) NS-WD and (b) DNS systems. $\Delta_{peak}$ is the fractional percentage deviation of the inferred value from the real peak. The optimum choice of prior is determined by minimizing $\chi^{2}$ and the fractional mean deviation from the input value. Priors (2) and (3) from \Fref{priors} perform best for DNS and NS-WD systems respectively.}
\label{Fig:perf}   
\end{figure*} % Fig:perf
\begin{figure*}[!th]
\subfigure[Neutron star - white dwarf systems]{\includegraphics[width=0.38 \textwidth,angle=90,trim=.5cm .5cm .5cm 1.9cm, clip=true]{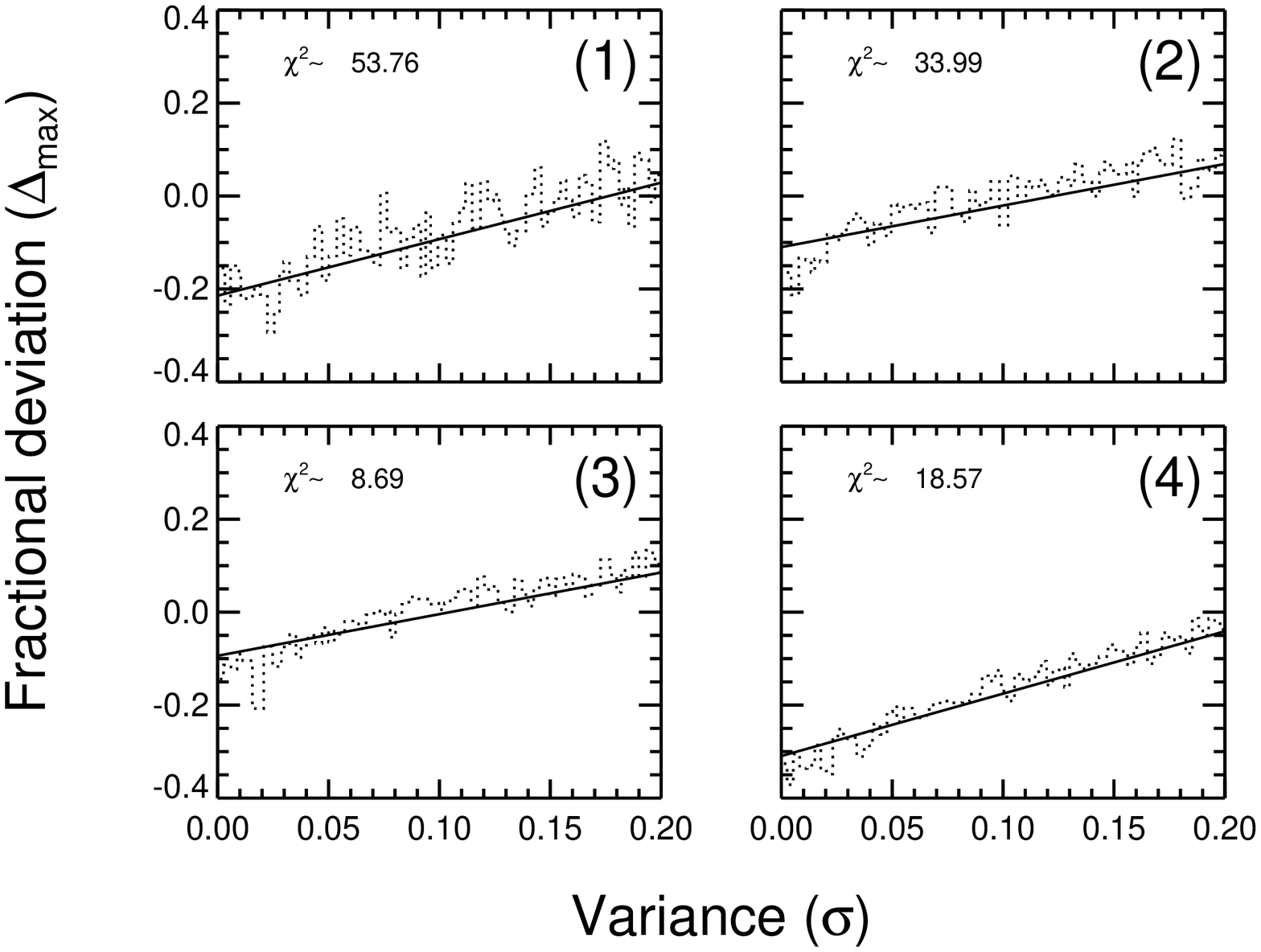} 
	 \label{Fig:maxW}}
 \subfigure[Double neutron stars systems]{\includegraphics[width=0.38 \textwidth,angle=90,trim=.5cm .5cm .5cm 1.9cm, clip=true]{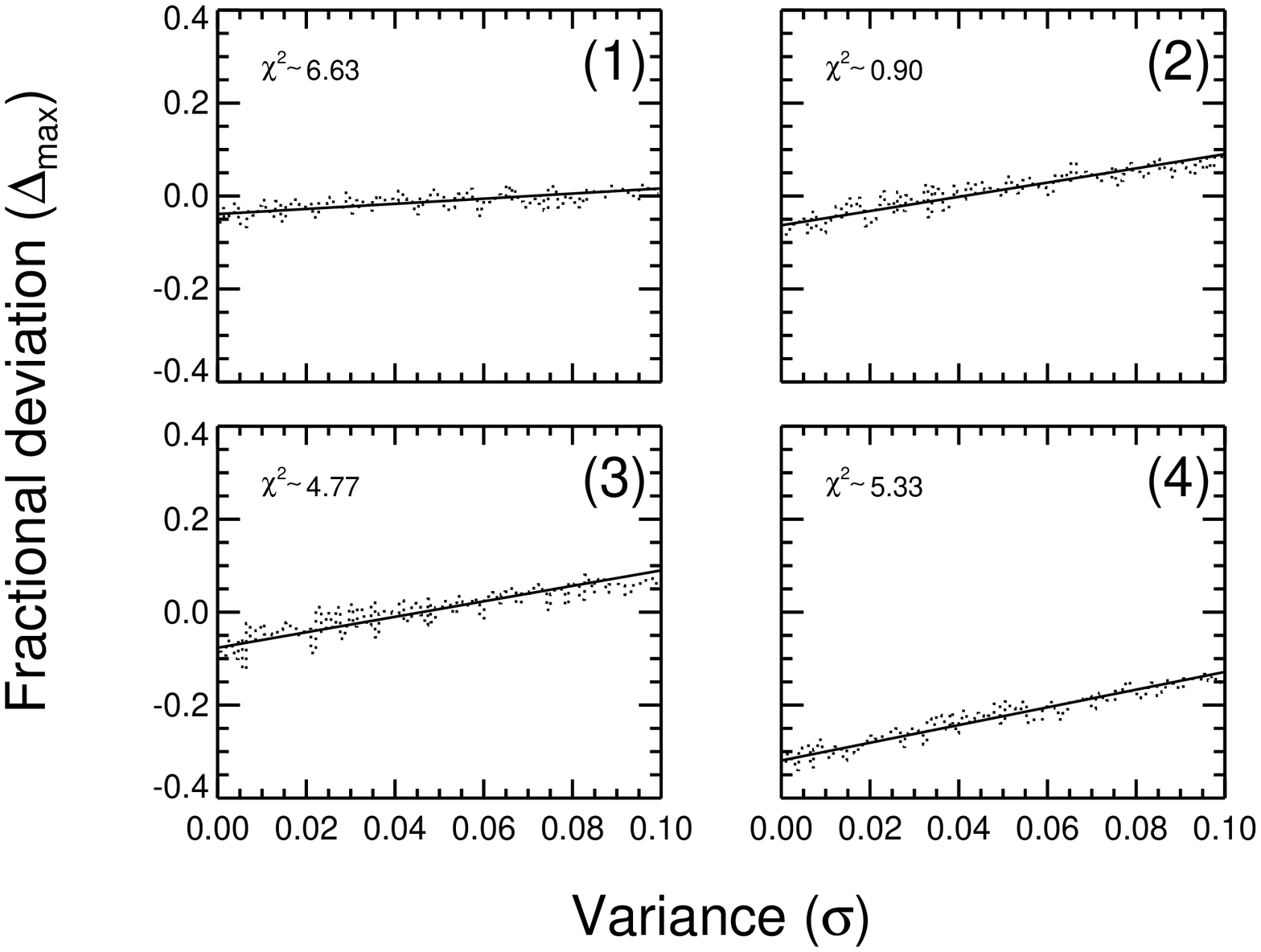} 
	 \label{Fig:maxD}}
 \caption{Performance of the inference algorithm for (a) NS-WD and (b) DNS systems. $\Delta_{max}$ is the fractional percentage deviation of the inferred value from the real maximum cut-off. The optimum choice of prior is determined by minimizing $\chi^{2}$ and the fractional mean deviation from the input value. Priors (2) and (3) from \Fref{priors} perform best for DNS and NS-WD systems respectively.}
 \label{Fig:max}   
\end{figure*}
 %Fig:max

The shape of a ``weakly informative'' prior will dynamically change with the data sample that is used for inference. For instance, for very tightly bound data sets such as DNS systems, a sharply peaked prior can qualify as weakly informative, while the same prior will certainly be more informative for a more dispersed data set such as NS-WD systems. Our goal is to quantitatively find an optimum choice rather than qualitatively assign priors.

Once we find optimum priors for DNS and NS-WD systems (see also \Sref{accu}), we use Monte-Carlo simulations to numerically estimate the accuracy level we can reach with our approach.

	\subsection{Accuracy and Error Estimation}\label{Sec:accu}

In order to test the performance of the algorithm that we use to infer the underlying  NS mass distribution, we run a series (i.e. $10^{5}$) of simulations. For each step, we construct distributions that have random peaks between $\mu\equiv$[0,3]$\msun $ with associated random variances between $\sigma\equiv$[0,0.3]. From these distributions, we randomly select the same number of samples as our ``observed'' sample. Next, we artificially corrupt these simulated samples by introducing random errors consistent with uncertainties in observations from the real data. We then use these smeared samples to test whether we can recover the ``input/known'' distribution.

To quantify performance we calculate the variance of two quantities. For each realization we compare the peaks and the maximum cutoff values of the input distribution and that of the predicted posterior we obtain with our algorithm. 

\Fref{perf} shows the fractional deviation from the input peak as a function of the intrinsic variance. The performance of the inference algorithm is quantified for NS-WD and DNS systems separately in \Fref{perfW} and \Fref{perfD}, respectively. The four panels illustrate the performance for the priors shown in \Fref{priors}. The numbers 1--4 on the upper right corner of each panel of \Fref{perf} refer to priors which incrementally widen into almost a flat uninformative prior (see \Fref{priors} for the corresponding hyper-parameters).

The performance is measured by the absolute value of $\Delta$ which is the fractional difference between the ``input'' and ``inferred'' value. The behavior of the variance $\chi^{2}$ is also used in conjunction as a measure of performance consistency. 

We repeat the same procedure in \Fref{max} to quantify performance for recovering the maximum cutoff value. 

Clearly shown by varying performances in \Fref{perf} and \Fref{max}, the priors that can be used to reliably infer the underlying distribution for DNS and NS-WD systems have to bear different characteristics. To infer the ``peak'' and the ``maximum cutoff value'' for DNS and NS-WD systems, priors (2) and (3) in \Fref{priors} perform best, respectively, by producing the most accurate inferences.

For simulated tight distributions similar to DNS systems, the mean accuracy of the inferred peak value is better than $99$\%. For wider distributions resembling NS-WD systems, the predicted peak has $>98$\% accuracy. Both for DNS and NS-WD systems, we find that the underlying real maximum cutoff value cannot be larger than $10\%$ of the inferred value. 

We also run a series of simulations in order to examine whether the algorithm can detect distributions that have been artificially skewed at varying levels and directions. Even at modest levels of input skewness, we find that the inferred shape (whether it exhibits skewness or not) is consistent with the underlying distribution with more than 99\% confidence (see \Sref{appr} and \Sref{maxx} for  discussion).

	\subsection{MCMC Algorithm Performance}\label{Sec:acce}

The technical core of Bayesian inference for models with analytically intractable posterior distributions is to use MCMC based algorithms. It is important to conduct detailed performance tests for the MCMC method used for inference in order to assess probabilistic biases and sources of sensitivity. Especially for sparse data, the probability distribution function (pdf), and hence the inferred confidence contours can be very sensitive to sampling steps. For instance, if the sampling acceptance rates for the M-H steps are not closely monitored and the inference process is not tuned for optimum sampling, over- or under-sampling, especially from the tails of the posterior distribution, will bias the inference. For sparse data, the introduced bias can potentially be very significant. 

It has been shown that for univariate target distributions the optimum acceptance rates for M-H based MCMC algorithms lie between 45--55\%, which 
may go down to as much as 23\% for multivariate target distributions \citep{Roberts:97}. For univariate distributions, acceptance rates higher than $60\%$ are indicative of over-sampling. Therefore, artificially high acceptance rates will result in underestimating the width of the distribution. On the other hand, acceptance rates below $30\%$ will over estimate the spread of the confidence contours by under sampling the parameter space \citep{Roberts:01}. We tune the M-H sampling steps to keep the acceptance rate in the optimum 45--55\% range.

Further quantitative insight into whether the MCMC sampler operates optimally can be gained by calculating the autocorrelation function for consecutively sampled model parameters. Each successive step will have a certain level of correlation, since they are simulated from a Markov chain. The extent of autocorrelation is an indication for the efficiency and, in general, performance of a particular MCMC algorithm.

\begin{figure}[]
\begin{center}
\includegraphics[width=0.37 \textwidth,angle=90,trim=.5cm .5cm .cm 1.2cm, clip=true]{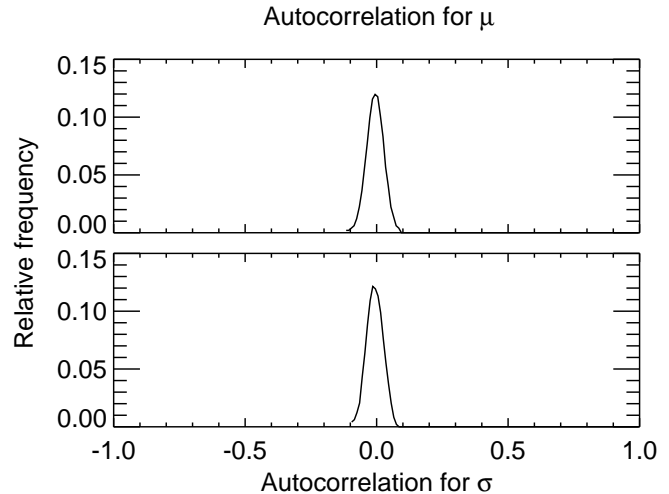} 	
\caption{Autocorrelation plots for consecutive MCMC steps sampling the parameter space of the mean neutron star mass $\mu$ and the Gaussian half width $\sigma$.}
\label{Fig:auto}   
\end{center}
\end{figure}

We calculate the autocorrelation function for both steps of the MCMC algorithm that sample the Gaussian mean $\mu$ and half-width 
$\sigma$ parameter. Figure~\ref{Fig:auto} shows that the steps in the MCMC sampling scheme for both $\mu$ and $\sigma$ have very small autocorrelation, which suggests that the MCMC algorithm efficiently explores the posterior distribution for the model parameters.

 %Fig:auto

\section{Summary}\label{Sec:summ}

We overview the physical processes that tune masses of NSs in \Sref{theo}. In order to theoretically estimate the viable range for NS masses, we derive the birth mass (\Sref{birth}, $M_{birth}=1.08$--$1.57\msun $) and the amount of mass expected to be transferred onto recycled NSs during the binary phase (\Sref{accr}, $\Delta m_{acc}\approx 0.1$--$0.2\msun$). We then discuss why the constraints on the maximum NS mass ($M_{max}=1.5$--$3.2\msun $) are less stringent and comment on the sources of uncertainties in \Sref{maxi}.

In order to maintain a uniform approach in our analysis, we refrained from including additional constraints that may arise from assumptions such as the possible relationship between the binary period and the mass of the remnant white dwarf (i.e., $P_{b}-m_{2}$ relationship) suggested by \cite{Rappaport:95}. While more elaborate and hierarchical implementation methods may be utilized in deducing ramifications of other assumptions, a use of more inclusive approaches may only convolute the mass inference, which is contrary to the goal of this work. Throughout our analysis, we only assume that Einstein's prescription for general relativity is correct and include mass measurements which are considered secure (\Sref{obs}).

In \Sref{evol} we point to the diversity of proposed evolutionary scenarios for DNS and NS-WD systems. Despite the limitations of constraints on their evolution, unlike isolated NSs, for pulsars in binary systems the precise measurements of the orbital parameters offer additional leverage to constrain the production channels.

We then subject the pulsar mass measurements to a detailed statistical analysis. In \Sref{esti} we show that a Bayesian method offers an effective means for inference. To alleviate the subjective nature, we use Markovian decision making algorithms in choosing the priors that produce the most accurate prediction for each sub-population. After we calculate the underlying NS mass distribution through posterior predictive densities,  in \Sref{algo} we use a large sample of simulated and artificially corrupted data to test the performance of this approach and quantify uncertainty. The width of the underlying distribution of NSs shown in \Fref{post} is then projected onto \Fref{psr} for visual reference.

\section{Discussion and Conclusions}\label{Sec:disc}

	\subsection{Previous Studies}\label{Sec:prev}

The first article that reviewed pulsar mass measurements in order to deduce the range of masses NSs can attain, was published by \cite{Joss:76}. They used mass measurements from 5 sources (PSR B1913+16, Her X-1, Cen X-3, SMC X-1, and 3U0900$-$40), which were predominantly X-ray sources, and found a marginally consistent range of $1.4$--$1.8\msun$. 

\cite{Finn:94} attempted to use Bayesian statistical techniques for the first time to infer limits on the NS mass distribution. By using the mass measurements of only 4 radio pulsars (PSRs B1913+16, B1534+12, B2127+11C, and B2303+46), he concluded that NS masses should fall mainly in the range between $1.3$--$1.6\msun$. The statistical approach he utilized did not, however, offer a means to measure the reliability and the predictive power. 

A comprehensive paper on pulsar masses was published by \cite{Thorsett:99}. Their analysis based on 26 sources yielded a remarkably tight mass range at $1.38^{+0.06}_{-0.10}\msun$. The width of their NS mass inference was mainly driven by the narrow error bands of the DNS mass measurements. 

The recent work by \cite{Schwab:10} suffers from other limitations. They analyze masses of 14 sources with an approach based on the comparison of the cumulative distribution function (CDF) with an idealized Gaussian. It is well understood that the K-S test should be used with caution in cases where deviations occur in the tails \citep{Mason:83}. Additionally, even in data samples where the number of outliers in the tails are considerably larger and associated measurement errors are taken into account, a K-S approach will still remain inadequate in quantifying the significance of the outliers. Therefore, while the bimodal feature found for the initial mass function (i.e. $M_{birth}$) may be consistent with theoretical expectations for remnant masses produced by electron-capture versus Fe-core collapse SNe \citep{Podsiadlowski:04}, the evidence for a deviation of $M_{birth}$ from a unimodal distribution is still tentative. In order to firmly establish a potential multi-modal feature for the NS birth mass distribution, a more diverse sample tested with more rigorous statistics are required.

	\subsection{Statistical Approach to Infer Underlying Distributions}\label{Sec:appr}

While the statistical model we use and the inference method we develop (\Sref{mode}) are specifically tailored for pulsar observations, it is detailed and modular enough that makes it a useful generic tool. There are several levels of challenges when adapting Bayesian methods:
(1) There is a wide misconception on how priors should be chosen for Bayesian inferences and the level of subjectivity they inject into the prediction. (2) Once an appropriate prior is chosen, the Monte-Carlo method used for parameter estimation has to follow Markovian steps rather than just random sampling. (3) Then, in order to prevent over- or under-sampling, the process has to be tuned for optimal sampling. This can be achieved by closely monitoring the acceptance rates of the MCMC steps. (4) Another important step for building a robust approach is to subject the algorithm to rigorous tests where simulated data with realistic errors and biases are used. (5) It is imperative to {\em quantify} the predictive power of the algorithm. Most convincingly, as demonstrated in \Fref{perf} and \Fref{max}, this can be accomplished by monitoring how much the inferred values deviate from simulated and corrupted input. In our case, we produce $10^{5}$ distributions with random peaks between $\mu\equiv$[0,3.0]$\msun$ and variances between $\sigma\equiv$[0,0.3] which are then used for the testing process. This procedure yields independent confidence estimates for the peak and maximum cutoff values.

The power of Bayesian inference is not in parameter estimation alone, but more in producing realistic predictions (see \Fref{post}). As demonstrated in \Sref{algo}, the reliability of Bayesian predictions can be quantified. We show that the method we use to infer the underlying NS mass distribution performs remarkably well. Even for limited samples, we achieve $\sim$99\% accuracy in predicting the peak and maximum cutoff values. In order to be conservative in our estimates, we run our testing procedure with 17 randomly chosen samples where, in fact, we have 18 well measured masses both for DNS and NS-WD systems.

Another parameter that we are interested in is the potential skewness of the underlying NS mass distribution. A universal EOS that consistently describes the micro-physics of NS matter will induce a truncation limit on the underlying distribution. Consequently, such a truncation limit, if it exists, will define the transition region at the high mass end where NSs are expected to collapse into stellar mass black holes --- this limit will delineate where stellar mass black holes form.

	\subsection{Maximum Mass Limit}\label{Sec:maxx}

We test whether the method is sensitive enough to detect signatures of a potential truncation, particularly at the high mass end of the underlying NS mass distribution. We find that even in cases where a mild truncation is imposed onto the input distribution, the algorithm produces results with $\sim$99\% consistency. Both predicted distributions for NSs in DNS and NS-WD systems in \Fref{post} are consistent with symmetric shapes (both with skewness parameter $|\gamma_{1}|<0.06$), and show no signs of deviation in favor of a truncation on either end.

This has important ramifications: The lack of truncation indicates that, in particular, the high mass end is driven by evolutionary constraints. Evolutionary processes such as long term stable accretion will naturally produce a wider distribution, while constraints due to general relativity or a universal EOS would produce a strict upper limit, which would manifest itself as a truncation limit. The lack of truncation in the shape of the underlying NS mass distribution, as a result, rules out the possibility that the upper mass limit is set by general relativity or the EOS. Therefore, the 2$\msun$ maximum mass limit implied by NS-WD systems should be considered as a minimum secure limit to the maximum NS mass rather than an absolute upper limit to NS masses.

	\subsection{Central Density and the Equation of State}

All EOSs that require a maximum NS mass $M_{max}\le2\msun$ are ruled out. The implied stiffness of the EOS largely precludes the presence of meson condensates and hyperons at supranuclear densities. Consequently, lower central densities, larger radii and thicker crusts for NSs are favored \citep{Shapiro:83}.

The energy density-radius relation implied by \cite{Tolman:39}, when combined with the causality limit, gives an analytical solution for an upper limit on the central density
\bee
\rho_{c}M^{2}=15.3 \times 10^{15} \msun^{2} \,\text{g}\,\text{cm}^{-3}.
\eee 
With a 2$\msun$ secure lower limit on the maximum NS mass, we set a 95\% confidence upper limit to the central density of NSs, which is 
\bee
\rho_{max}<3.83\times10^{15} \text{g}\,\text{cm}^{-3}
\eee 
corresponding to $\approx$11$\rho_{s}$ for a fiducial saturation threshold $n_{s}\sim 0.16$ fm$^{-3}$.

Exotic matter such as hyperons and Bose condensates significantly reduce the maximum mass of NSs. Therefore, a strict lower limit on the maximum NS mass M$_{max}>2\msun$ rules out soft EOSs with extreme low density softening %(e.g., GS1, GS3)
and require the existence of exotic hadronic matter \citep[see][for review]{Lattimer:07}. NSs with deconfined strange quark matter mostly have maximum predicted masses lower than 2$\msun$. Hence, EOSs with strange quark matter that predict maximum masses smaller than 2$\msun$ %(e.g., SQM1 and SQM3) 
can also be ruled out as viable configurations for NS matter.

	\subsection{Evidence for Alternative Evolution and the Formation of Massive Neutron Stars?}

A 2$\msun$ upper limit to masses of NSs in NS-WD system poses a problem. If all millisecond pulsars were indeed NSs that are recycled from a first generation of normal pulsars, the implied distribution should be consistent with a recycled version of the initial mass distribution. While the peaks of the distributions for DNS and NS-WD systems are consistent with the expectations of standard recycling (\Sref{accr}), the widths imply otherwise. As shown in \Fref{post}, $\Delta m_{acc}=0.15\msun$ lies perfectly within the expected range. However, with typical accretion rates experienced during the LMXB phase ($\dot{m}_{acc} \sim 10^{-3}\,\dot{\text{M}}_{\text{Edd}}$), NSs with masses $\sim$2$\msun$ such as PSR J1614$-$2230 cannot be formed. Even with initial masses of $\sim$1.6$\msun$ these sources need to accrete $\Delta m \approx0.4\msun$ during their active accretion phase. This requires long term stable active accretion at unusually high rates.

Based on the \ppd demographics of millisecond pulsars, \cite{Kiziltan:09} argue that $\approx$ 30\% of the millisecond pulsar population may be produced via a non-standard evolutionary channel. This prediction falls in line with a distribution that has a consistent recycled peak but has an unusual width which extends up to 2$\msun$. While it is difficult to quantify the formation rate(s) of non-standard processes that may produce these NSs, it is clear that the standard scenario requires at least a revision. Such a revision should consistently reconcile for the observed \ppd distribution of millisecond pulsars, along with the long term sustainability of unusually high accretion rates that is required to produce the second generation of massive NSs. 

The only viable alternative to a major revision of the mass evolution implied by the standard recycling scenario, also corroborated by the lack of truncation of the underlying NS mass distribution, is then to form massive NSs.

\acknowledgements 
The authors thank P. Freire for sharing updated pdfs from which some of the NS mass estimates were extracted in \Tref{nswd}. B.K. thanks the ITC at Harvard-Smithsonian Center for Astrophysics, CIERA at Northwestern University, Department of Physics at West Virginia University for their hospitality; and Jonathan E. Grindlay, Saul Rappaport, Deepto Chakrabarty, Duncan Lorimer for stimulating discussions. B.K. and S.E.T. acknowledge NSF grant AST-0506453.

%%% Bibliography
\bibliographystyle{apj}	%(uses file "apj.bst")

%%% appendix 
\appendix
\phantomsection \label{Sec:appendix}

\section{Inference under the Bayesian Model for the Neutron Star Mass Distribution}\label{Sec:model}

	\subsection{Model Formulation}

Based on the statistical model formulation for the NS mass distribution developed in Section 
\ref{Sec:mode}, the 
%
%We analytically calculate the underlying neutron star mass distribution implied by the observed mass estimates. 
%We perform our calculations for
%\[
%m_{i}=\m_{i}+w_{i},\, i=1,...,n
%\]
%where $m$ represents pulsar mass estimates, $\m$ is the underlying ``real'' mass distribution 
%with associated $w_{i}$ errors 
%that gives an observed mass estimate $m_{i}$ for each observation ``$i$''. $\m$ is modeled as a normal distribution 
%$N(\m; \mu, \sigma^{2})$ independent of $m$ and $w$. The error variances $S_{i}^{2}$ for $w_{i}$ are obtained from 
%the error bands where $w_{i}\sim N(0,S_{i}^{2})$.
%
%The proposed model yields
%\[
%f(\m,w) = N(\m;\mu,\sigma^{2})\,N(\m;0,S^{2})
%\]
%which, through transformation, implies a $N(m;\mu,\sigma^{2}+S^{2})$ distribution for $m=\m+w$.
%
%Therefore, the 
%
likelihood for the data \{$(m_{i}, S_{i}):i=1,...,n$\} is given by
\[
\mathcal{L}(\mu,\sigma^{2};\text{data}) = \prod^{n}_{i=1}\left [2\pi (\sigma^{2}+S_{i}^{2})  \right]^{-1/2} \,
e^{-\frac{(m_{i}-\mu)^{2}}{2(\sigma^{2}+S_{i}^{2})}}	 \propto 
\left\{ \prod^{n}_{i=1} (\sigma^{2}+S_{i}^{2})^{-1/2} \right\}
\exp\left\{ -\frac{1}{2} \sum_{i=1}^{n} \frac{(m_{i}-\mu)^{2}}{(\sigma^{2}+S_{i}^{2})} \right\}
\eqref{Eq:like}
\]

The Bayesian model is completed with independent normal $N(a,b^{2})$ and inverse-gamma 
$IG(c,d)$ priors for $\mu$ and $\sigma^{2}$, respectively. Specifically,
\bee
\pi(\mu) = (2\pi b^{2})^{-1/2} \exp\left[- \frac{(\mu-a)^{2}}{2b^{2}} \right]
\label{Eq:pri1}
\eee
and
\bee
\pi(\sigma^{2}) = \frac{d^{c} \exp(-\frac{d}{\sigma^{2}})}{\Gamma(c)\sigma^{2\,(c+1)}}
\label{Eq:pri2}
\eee
for fixed hyper-parameters $(a,b,c,d)$. Note that the prior mean for $\sigma^{2}$ is given by $d/(c-1)$
(provided $c > 1$).

Combining the likelihood with the priors for $\mu$ and $\sigma^{2}$, the posterior distribution for the model parameters can be written proportional to 
\bee
p(\mu,\sigma^{2} \mid \text{data})  & \propto &	\pi(\mu) \pi(\sigma^{2}) \mathcal{L}(\mu,\sigma^{2};\text{data})
\nonumber\\
& \propto & \exp\left[-\frac{(\mu-a)^{2}}{2b^{2}}\right] \frac{\exp(-\frac{d}{\sigma^{2}})}{\sigma^{2\,(c+1)}} \times 
\left\{ \prod^{n}_{i=1} (\sigma^{2}+S_{i}^{2})^{-1/2} \right\}
\exp\left\{ -\frac{1}{2} \sum_{i=1}^{n} \frac{(m_{i}-\mu)^{2}}{(\sigma^{2}+S_{i}^{2})} \right\}. \nonumber
\eee
As discussed in Section \ref{Sec:mode}, although the normalizing constant for the posterior density is not available in closed form, we can utilize MCMC sampling from $p(\mu,\sigma^{2} \mid \text{data})$, which results in full inference of the model parameters, as well as the posterior prediction of the NS mass distribution.

	\subsection{MCMC Posterior Simulation Method}\label{post}

The MCMC algorithm updates dynamically the two parameters, $\mu$ and $\sigma^{2}$, by sampling
from their posterior full conditional distributions. After convergence, the resulting samples are 
realizations from the posterior distribution $p(\mu,\sigma^{2} \mid \text{data})$. 

The posterior full conditional distribution for $\mu$ is proportional to 
\[
p(\mu \mid \sigma^{2},\text{data}) \propto \exp\left[-\frac{(\mu-a)^{2}}{2b^{2}}\right] 
\exp\left\{ -\frac{1}{2} \sum_{i=1}^{n} \frac{(m_{i}-\mu)^{2}}{(\sigma^{2}+S_{i}^{2})} \right\}
\]
an expression which can be completed to a normal distribution with mean
\[
\tilde{a} = 
\frac{a+b^{2}\,\sum_{i=1}^{n}m_{i}(\sigma^{2}+S_{i}^{2})^{-1} }{1+b^{2} \sum^{n}_{i=1}(\sigma^{2}+S_{i}^{2})^{-1}}
\] 
and variance 
\[
\tilde{b}^{2}= \frac{b^{2}}{1+b^{2} \sum^{n}_{i=1}(\sigma^{2}+S_{i}^{2})^{-1}}.
\]    
%
%\[
%\tilde{b}^{2} = \frac{b^{2}}{1+b^{2} \sum^{n}_{i=1}(\sigma^{2}+S_{i}^{2})^{-1}}.
%\]

However, the posterior full conditional for $\sigma^{2}$,
\[
p(\sigma^{2} \mid \mu,\text{data})  \propto p^{*}(\sigma^{2}) = 
\frac{\exp(-\frac{d}{\sigma^{2}})}{\sigma^{2\,(c+1)}} 
\left\{ \prod^{n}_{i=1} (\sigma^{2}+S_{i}^{2})^{-1/2} \right\}
\exp\left\{ -\frac{1}{2} \sum_{i=1}^{n} \frac{(m_{i}-\mu)^{2}}{(\sigma^{2}+S_{i}^{2})} \right\},
\]
can not be recognized as a standard distributional form, and we thus use a Metropolis-Hasting (M-H) step to update $\sigma^{2}$ given $\mu$. For the M-H proposal distribution, we use the full conditional for $\sigma^{2}$ from the special case of the model that does not include errors in measurement for the pulsar mass estimates (i.e. $S_{i}^{2} \equiv 0$, for all $i$). %as a standard distributional form, and we thus use a Metropolis-Hasting (M-H) step to update $\sigma^{2}$ given $\mu$ where $S_{i}^{2} \equiv 0$, for all $i$. %For the M-H proposal distribution, we use the full conditional for $\sigma^{2}$ from the special case of the model that does not incorporate errors in measurement for the  pulsar mass estimates (that is, $S_{i}^{2} \equiv 0$, for all $i$). 
Under this simplified version of the model, $\sigma^{2}$ has an inverse-gamma $IG(\tilde{c}, \tilde{d} )$ posterior full conditional distribution, where
$\tilde{c}=$ $0.5n + c$ and $\tilde{d}=$ $d + 0.5 \sum_{i=1}^{n}(m_{i}-\mu)^{2}$.

Denoting by $\sigma^{2 (t)}$ the current state of the Markov chain for $\sigma^{2}$, the 
M-H step proceeds as follows: We draw a proposed value $\tilde{\sigma}^{2}$ from the 
$IG(\tilde{c},\tilde{d})$ distribution, and then compute the acceptance probability 
\[
q = \min\left\{ 1, \frac{p^{*}(\tilde{\sigma}^{2})}{p^{*}(\sigma^{2 (t)})} \times 
\frac{g(\sigma^{2 (t)})}{g(\tilde{\sigma}^{2})}  \right\}
\]
where
\[
g(u)= \tilde{d}^{\tilde{c}}\frac{\exp(-\tilde{d}/u)}{\Gamma(\tilde{c}) u^{\tilde{c}+1}}
\] 
denotes the density of the $IG(\tilde{c},\tilde{d})$ distribution, and as indicated 
above, $p^{*}(\sigma^{2})$ is the density of the un-normalized posterior full conditional 
distribution for $\sigma^{2}$. Then, we obtain the new state of the chain for $\sigma^{2}$ 
through
\[
\sigma^{2 (t+1)} = \left\{ 
\begin{array}{l l}
  \tilde{\sigma}^{2} & \quad \text{with probability $q$}\\
  \sigma^{2 (t)} & \quad \text{with probability 1-$q$}\\
\end{array} \right.
\]
that is, a {\em stochastic rejection} step to determine whether the proposed value 
$\tilde{\sigma}^{2}$ is accepted.

Therefore, the MCMC method to sample from the posterior distribution $p(\mu,\sigma^{2} \mid \text{data})$ involves overall the following iterative procedure:
\begin{itemize}
\item Start with initial values $(\mu^{(0)},\sigma^{2 (0)})$.
\item If the current iteration is ($\mu^{(t)}, \sigma^{2 (t)}$), obtain the next 
iteration ($\mu^{(t+1)}, \sigma^{2 (t+1)}$) through the following two updates: 
\begin{itemize}
\item Draw $\mu^{(t+1)}$ from $N(\tilde{a},\tilde{b}^{2})$, where $\tilde{a}$ and 
$\tilde{b}^{2}$ are computed using $\sigma^{2} \equiv \sigma^{2 (t)}$.
\item Draw $\sigma^{2 (t+1)}$ using the M-H step, where both $p^{*}(\cdot)$ 
and $g(\cdot)$ are evaluated using $\mu \equiv \mu^{(t+1)}$.
\end{itemize}
\end{itemize}

\section{Posterior Predictive Distribution}\label{Sec:post}

%The resulting posterior samples $\{(\mu^{(l)}, \sigma^{2 (l)}): l=1,...,L\}$ can be used for 
%inference with ``$l$'' M-H iteration steps. 
%Ideally, the first 5-10\% of the M-H steps will be considered the ``burn-in'' cycle and, 
%along with the acceptance ratio, should be 
%monitored closely in order to prevent over or under-sampling.
%

The posterior samples from $p(\mu,\sigma^{2} \mid \text{data})$ can be used
to estimate the density of the NS mass distribution. The formal Bayesian 
estimate is given by the posterior predictive density 
$\mathcal{P}(\m_{0} \mid \text{data})$ (which is the posterior density
given the observed data), for a ``new'' unobserved pulsar with unknown mass 
$\mathcal{\m}_{0}$. Hence, $\mathcal{\m}_{0}$ is a parameter that we seek to 
estimate under the full Bayesian model that includes also the NS mass 
distribution parameters $\mu$ and $\sigma^{2}$. 

Based on the model for the 
observed mass estimates developed in Section \ref{Sec:mode}, the augmented 
model that incorporates $\mathcal{\m}_{0}$ involves two new terms: a 
$N(\mathcal{\m}_{0};\mu,\sigma^{2})$ distribution for the unknown mass 
$\mathcal{\m}_{0}$, and a $N(w_{0};0,S_{0}^{2})$ component for the error that 
would arise if we were to observe the pulsar that has a mass estimate $m_{0}$ associated
with $\mathcal{\m}_{0}$. Now, this model can be marginalized over $w_{0}$ to 
obtain the joint posterior distribution for $\mathcal{\m}_{0}$ and
$(\mu,\sigma^{2})$,
\[
p(\mathcal{\m}_{0},\mu,\sigma^{2} \mid \text{data}) \propto 
N(\mathcal{\m}_{0};\mu,\sigma^{2}) 
\left\{ \prod_{i=1}^{n} N(m_{i};\mu,\sigma^{2}+S_{i}^{2}) \right\} 
\pi(\mu) \pi(\sigma^{2}) 
= N(\mathcal{\m}_{0};\mu,\sigma^{2}) p(\mu,\sigma^{2} \mid \text{data}).
\]
Finally, $\mathcal{P}(\m_{0} \mid \text{data})$ is obtained by marginalizing 
the joint posterior $p(\mathcal{\m}_{0},\mu,\sigma^{2} \mid \text{data})$
over $\mu$ and $\sigma^{2}$:
\[
\mathcal{P}(\m_{0} \mid \text{data}) = \int\int N(\mathcal{\m}_{0};\mu,\sigma^{2}) \, 
p(\mu,\sigma^{2} \mid \text{data}) \,d\mu \,d\sigma^{2}
\eqref{Eq:prob}.
\]
Note therefore that the Bayesian estimate, $\mathcal{P}(\m_{0} \mid \text{data})$,
for the NS mass density {\em incorporates uncertainty} for parameters $(\mu,\sigma^{2})$ 
by averaging over their posterior distribution.
This can be contrasted with the predictive density under the likelihood 
approach, where the $N(\m_{0};\mu,\sigma^{2})$ NS mass density would simply
be estimated by replacing the parameter vector $(\mu,\sigma^{2})$ with its
maximum likelihood estimate. 

The posterior predictive density $\mathcal{P}(\m_{0} \mid \text{data})$
can then be readily estimated through straightforward Monte Carlo integration
suggested by Equation \ref{Eq:prob}, using the MCMC samples from the 
posterior distribution $p(\mu,\sigma^{2} \mid \text{data})$.

\end{document}